\documentclass[
hidelinks,
twocolumn,
secnumarabic,
amssymb, 
nobibnotes, 
aps, 
prb]{revtex4-2}
\usepackage{graphicx}
\usepackage{multirow}
\usepackage{lipsum} 
\usepackage[breaklinks=true]{hyperref} 
\usepackage[T1]{fontenc}
\usepackage[utf8]{inputenc}
\usepackage{textcomp} 
\usepackage{lmodern}
\graphicspath{{./figures/}}

\usepackage{dcolumn}
\newcolumntype{d}[1]{D{.}{.}{#1}} 
\usepackage{bm}
\usepackage{textgreek}
\usepackage{pslatex}
\usepackage{breakurl} 
\usepackage{booktabs}
\usepackage{tabularx}

\usepackage{comment}

\usepackage[colorinlistoftodos]{todonotes}

\usepackage[colorinlistoftodos]{todonotes}

\usepackage[colorinlistoftodos]{todonotes}

\usepackage{xcolor}
\hypersetup{
    breaklinks = true,
    colorlinks,
    pdftitle = {}, 
    pdfauthor = {},
    pdfkeywords = {},
    linkcolor={blue!80!black},
    citecolor={blue!80!black},
    urlcolor={blue!80!black},
    filecolor = black
}

\graphicspath{{./figures/}}

\newcommand{\feco}{Fe$_{\mathrm{0.75}}$Co$_{\mathrm{0.25}}$}

\begin{document}
\begin{sloppypar} 

\title{Perpendicular magnetic anisotropy in 
Au/FeCo/Au 
ultrathin films:\\
Combined experimental and first-principles study
}

\author{Justyn Snarski-Adamski}
\author{Hubert Głowiński}
\author{Justyna Rychły-Gruszecka}
\author{Piotr Kuświk}
\author{Mirosław Werwiński}
\email[Corresponding author: Mirosław Werwiński\\Email address: ]{werwinski@ifmpan.poznan.pl}

\affiliation{
Institute of Molecular Physics, Polish Academy of Sciences, 
Mariana Smoluchowskiego 17, 60-179 Poznań, Poland}

\date{\today}


\begin{abstract}

%
Magnetic tunnel junctions with a magnetic layer with perpendicular anisotropy are currently used in computer memories that do not require voltage sustaining.
An example of layers with perpendicular magnetic anisotropy are ultrathin FeCo films on Au substrate.
Here, we present an experimental and computational study of \feco{} layers with a thickness up to two nanometers.
The investigated FeCo layers are surrounded on both sides by Au layers.
%
%
In experiment, we found perpendicular magnetic anisotropy in FeCo polycrystalline films with a thickness below 0.74~nm (five atomic monolayers).
We also measured a strong surface contribution to effective magnetic anisotropy.
%
%
From the density functional theory we determined structural, electronic, and magnetic properties of the FeCo films.
The calculations showed an irregular dependence of magnetic anisotropy on FeCo layer thickness.
We observed perpendicular anisotropy for one- and three-atom FeCo monolayers and in-plane anisotropy for two- and four-atom monolayers.
However, the determination of running averages of magnetic anisotropy (of three successive thicknesses of atomic monolayers) leads to a close linear thickness dependence of magnetic anisotropy, similar to the experimental result.
%
%
The reason why the properties of the several-atom-thick FeCo layers differ so significantly from those of the bulk parent material is the disappearance of the central region of the layer whose properties approximate those of the bulk, and instead the existence of near-interface regions with properties altered by the presence of discontinuities.

\end{abstract}

\maketitle

\section{Introduction}

%
Spin transfer torque magnetoresistance random access memory (STT-MRAM) uses a perpendicular-anisotropy magnetic tunnel junction~\cite{huai_high_2018,ikeda_perpendicular-anisotropy_2010}.
In the junction, the ultrathin Fe, FeCo, and FeCoB magnetic films can be used as layers with perpendicular anisotropy,
and
the magnetization direction of the layer can be switched with assistance of a bias voltage~\cite{
ikeda_perpendicular-anisotropy_2010,
maruyama_large_2009,
shiota_voltage-assisted_2009,
nozaki_voltage-induced_2010,
shiota_induction_2012}.
%
%
For Fe and FeCo layers, perpendicular magnetic anisotropy (PMA) occurs at a thickness of a few atomic monolayers~\cite{gay_spin_1986, liu_perpendicular_1990}.

%
%
%
By convention, the thin film can be divided into regions of lower and upper interface and volume center.
For films with a thickness of only a few atomic monolayers, such an approach is no longer justified, since the central volume part has disappeared and the layer is composed of only near-surface regions.
Moreover, in idealized models with atomically sharp interfaces, when we change the layer thickness by one atomic monolayer, say from two to three monolayers, it is a significant qualitative change.

%
Combining experiment and calculations, we will determine the effect of the layer thickness on the properties of the FeCo layer in the Au/\feco{}/Au heterostructure.
Similar systems, Au/Fe$_{0.8}$Co$_{0.2}$/MgO and multilayer [Fe$_{0.3}$Co$_{0.7}$/Au]$_{\times{}20}$, exhibit PMA in the FeCo layer of less than four and six atomic monolayers, respectively~\cite{shiota_voltage-assisted_2009, wang_magnetic_2011}.
As we can notice, PMA can be found in the layers with a wide range of compositions, including Fe$_{0.3}$Co$_{0.7}$, Fe$_{0.8}$Co$_{0.2}$, and pure Fe.
We chose the Fe$_{0.75}$Co$_{0.25}$ alloy for its high magnetization and low intrinsic damping~\cite{schoen_ultra-low_2016, glowinski_influence_2019}.
Whereas, we selected the Au substrate because of the good match between fcc Au(001) and bcc FeCo(001) lattice parameters (a difference of about 1\%).

We will present experimental results for thickness gradient of Fe$_{0.75}$Co$_{0.25}$ film in Ti~3~nm\,/\,Au~60~nm\,/\,Fe$_{0.75}$Co$_{0.25}$~(0--2~nm)\,/\,Au~3~nm sample, followed by atomistic computational analysis (DFT). 
The theoretical results will help to interpret the experiment and expose the intrinsic properties of the multilayer.

%
%
%
We have prepared a series of Au/FeCo/Au models with FeCo layer thicknesses ranging from 1 to 15 atomic monolayers (up to 2.1~nm) and layer composition identical to the \feco{} alloy used in the experiment.
Our DFT study continue the tradition of calculating magnetic anisotropy in thin films, started in the 1980s with the prediction of perpendicular magnetic anisotropy in a single atomic monolayer of Fe~\cite{gay_spin_1986}.
Since the calculation of magnetic anisotropy for heterostructures takes a long time, calculations of thickness dependence are rarely performed.
For heterostructures with FeCo layer, previous DFT studies have usually assumed a fixed FeCo film thickness, such as three~\cite{li_thermally_2015, ong_strain_2015, ong_electric-field-driven_2016} or nine~\cite{peng_origin_2015,peng_giant_2017} atomic monolayers.
The mentioned models further imply an alternating ordering of Fe and Co monolayers~\cite{li_thermally_2015, ong_strain_2015, ong_electric-field-driven_2016,peng_origin_2015,peng_giant_2017}, leading to an artificial contribution to magnetic anisotropy.
In contrast, another study that indeed analyzed the dependence of magnetic anisotropy on FeCo layer thickness (in the Cu/FeCo/MgO/FeCo/Cu system) used the atomic sphere approximation (ASA)~\cite{zhang_perpendicular_2014}, which is too crude to enable the determination of indicative values of magnetic anisotropy energy~\cite{edstrom_magnetic_2015, hedlund_magnetic_2017}.

%
In this study, we performed the calculations in full-potential approach.
Furthermore, to model FeCo alloy, we used a virtual crystal approximation, which has proven itself in determining correct MAE trends and as a good starting point for more accurate calculations~\cite{burkert_giant_2004, diaz-ortiz_structure_2006}.
Since our computational model assumes a well-oriented Au(001) substrate with an epitaxial FeCo(001) layer, it is closer in geometry to the systems investigated by Shiota~\textit{et al.}~\cite{shiota_voltage-assisted_2009,shiota_induction_2012} than to the polycrystalline structure obtained in our experiment.

\section{Experimental details}

%
\begin{figure}[!ht]
\centering
\includegraphics[trim = 0 0 0 3,clip,width=0.47\columnwidth]{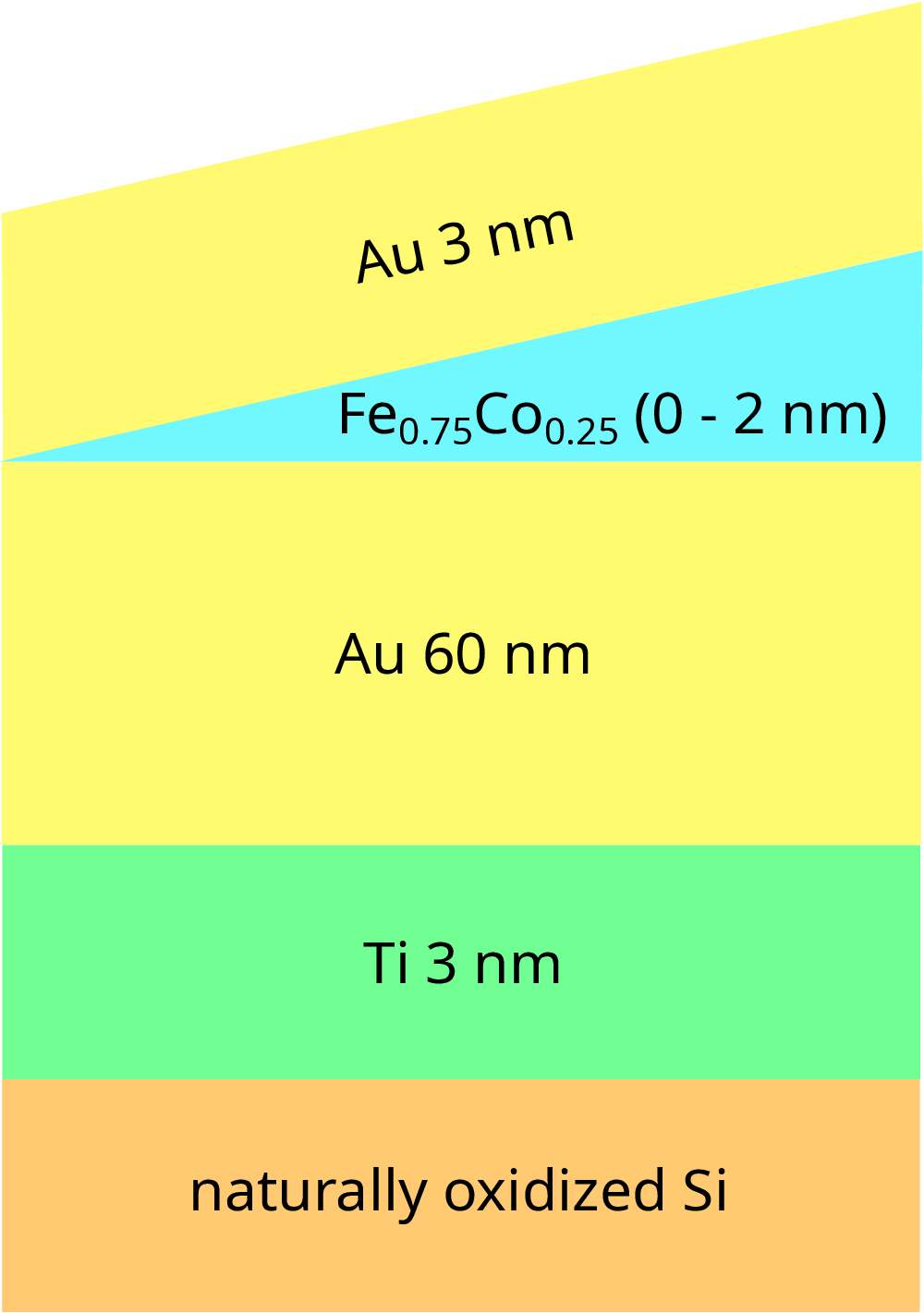}
\caption{
The arrangement of layers in the experimental heterostructure: Ti~3~nm\,/\,Au~60~nm\,/\,Fe$_{0.75}$Co$_{0.25}$~(0--2~nm)\,/\,Au~3~nm.
}
\label{fig:expt_heterostructure} 
\end{figure}

%
Using magnetron sputtering with alloy targets, we deposited a wedged-shaped film of Fe$_{0.75}$Co$_{0.25}$ with a thickness of 0 to 2~nm along a 20~mm-long sample of Ti~3~nm/Au~60~nm, and capped it with a layer of Au 3~nm, see Fig.~\ref{fig:expt_heterostructure}.
As a substrate, we used naturally oxidized silicon. 
We prepared the samples in Ar at a pressure of $1 \times 10^{-4}$~mbar. 
Whereas the base pressure was less than $5 \times 10^{-8}$~mbar. 
The deposition rates for all materials were calculated using data from a calibration sample measured with a profilometer.

%
We utilized a ferromagnetic resonance with a vector network analyzer (VNA-FMR) to
measure the sample locally, placing a 450~micron wide coplanar waveguide at various positions above the FeCo thickness gradient and sweeping an external magnetic field at a fixed frequency up to 40~GHz. 
In the out-of-plane configuration, we could measure ferromagnetic resonance only over a limited range of thicknesses and frequencies -- where the effective magnetization was low enough to saturate the sample.

%
We measured the hysteresis loops along the FeCo thickness gradient using the polar magneto-optical Kerr effect (pMOKE).
We determined the effective magnetic anisotropy constant from the equation $K_{\rm{eff}} = \frac{-H_a M_{\rm{S}}}{2}$, where $M_{\rm{S}}$ is the saturation magnetization and $H_a$ is the anisotropy field.
%
%
We determined the volume ($K_{\rm{V}}$) and surface ($K_{\rm{S}}$) contributions to magnetic anisotropy from the formula $K_{\rm{eff}}\,t = K_{\rm{V}}\,t + 2K_{\rm{S}}$.
Since the FeCo film has two interfaces, we represented its surface magnetic anisotropy as $2K_{\rm{S}}$.

We conducted all the measurements at room temperature.

\section{Computational details}

%
\begin{figure}[ht]
\centering
\includegraphics[clip,width=1.05\columnwidth]{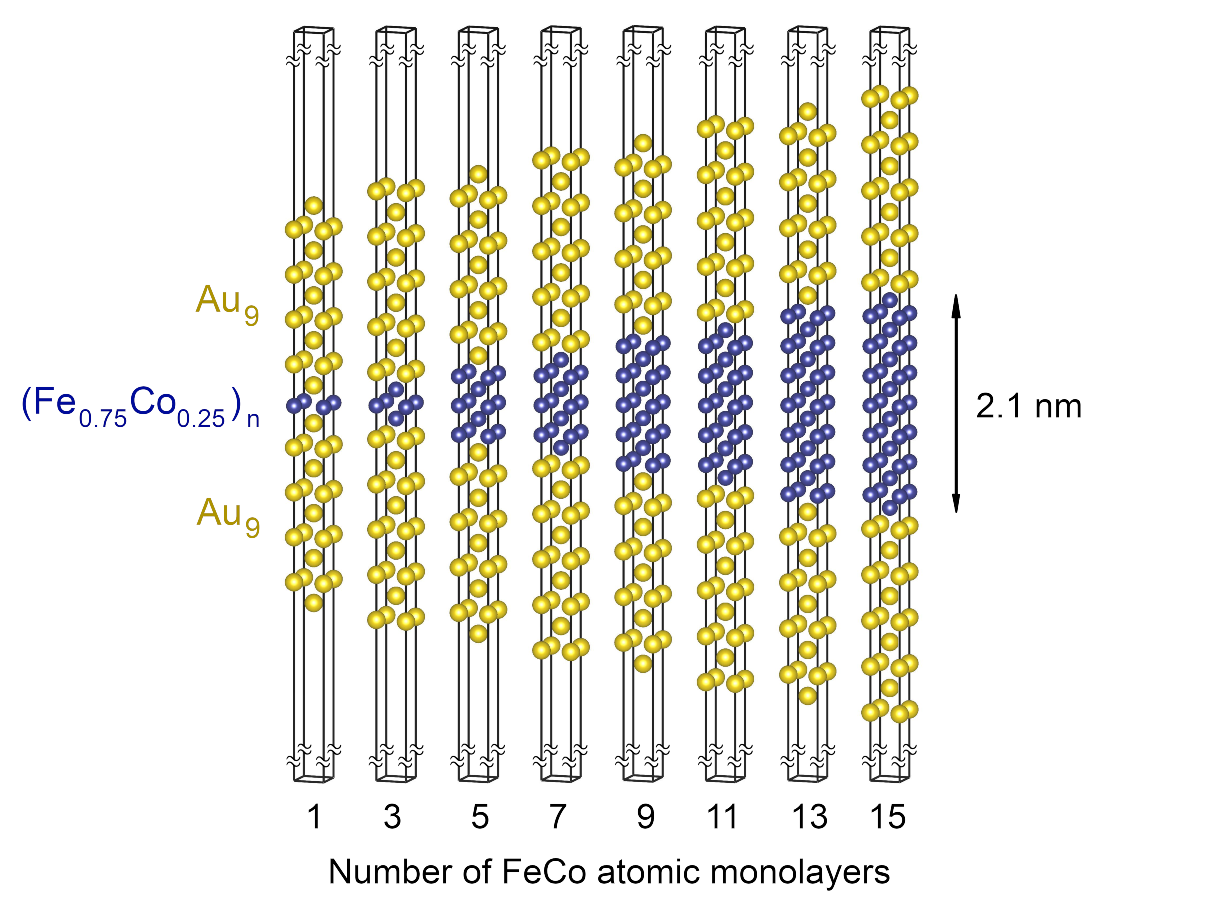}
\caption{ 
Unit cells of Au$_{9}$/(Fe$_{0.75}$Co$_{0.25}$)$_n$/Au$_{9}$ heterostructures.
Yellow balls represent Au atoms, and blue balls indicate Fe/Co atoms.
The thickness of FeCo layer varies from one to fifteen atomic monolayers (2.1~nm).
Although not shown, systems with an even number of FeCo monolayers have also been considered.
We applied the virtual crystal approximation to the atoms in the Fe$_{0.75}$Co$_{0.25}$ layer.
We placed nine Au atomic monolayers each above and below the FeCo layer.
The multilayer is surrounded by a vacuum with a thickness of no less than 50~\AA{}.
}
\label{fig:aufecoau_structs} 
\end{figure}

%
In order to model the Au/FeCo/Au multilayer in the range of experimental FeCo layer thicknesses (up to 2.1~nm), we prepared a series of Au/FeCo/Au heterostructures with FeCo layer thicknesses from 1 to 15 atomic monolayers and with a top and bottom Au layer thickness of 9 atomic monolayers, see Fig.~\ref{fig:aufecoau_structs}.
The chosen thickness of Au layers is a compromise between computational cost and sufficient reproduction of their properties.
We also included a vacuum of no less than 50~\AA{} in the $z$ direction of the unit cell.
In our models, we assumed that the bcc FeCo(001) layer ($a_\mathrm{bcc FeCo} = 2.84$~\AA{}) grows epitaxially on the fcc Au(001) substrate ($a_\mathrm{fcc Au} = 4.08$~\AA{}).
We matched the layers by combining the $1\times1$ bcc FeCo unit cell with $(1/\sqrt{2}) \times (1/\sqrt{2})$ fcc Au unit cell (tetragonal representation), see for example Ref.~\cite{xie_origin_2008}.
The lattice parameter of the $(1/\sqrt{2}) \times (1/\sqrt{2})$ Au unit cell is $4.08/\sqrt{2} \approx 2.88$~(\AA{}) and 
we took this value as a fixed in-plane lattice parameter for all the models.
It differs by about 1\% from the lattice parameter of bulk bcc \feco{}, which means a good alignment between the films.
However, recent studies for ultrathin Fe films indicated the appearance of
 boundary-induced body-centered tetragonal phase~\cite{ravensburg_boundary-induced_2024}, which may also affect the properties of the thinnest \feco{} films.
With a fixed lattice parameter $a$, we optimized the atomic positions in the $z$-direction using the spin-polarized scalar-relativistic approach with a 10$^{-3}$\,eV\,\AA{}$^{-1}$ force convergence criterion.

%
%
To solve the Kohn-Sham equations for a series of models, we applied the full-potential local-orbital code FPLO18.00-52~\cite{koepernik_full-potential_1999}.
We used generalized gradient approximation (GGA) in the form of the Perdew, Burke and Ernzerhof (PBE) functional~\cite{perdew_generalized_1996}.
%
%
We used the virtual crystal approximation to model the chemical composition of the FeCo layer.
This was possible because Fe and Co have consecutive atomic numbers (26 and 27).
Thus, for each atom in the \feco{} layer, we assumed a fractional atomic number of 26.25. 
%
%
Furthermore, we determined magnetic anisotropy energies (MAEs) as the difference in fully relativistic energies calculated for quantization axes oriented in the plane of the film and out of plane~\cite{opahle_full-potential_1999}.
To determine fully relativistic energies, we calculated a single fully relativistic iteration after self-consistent scalar-relativistic calculations.
We conducted the latter using a dense k-mesh of $60 \times 60 \times 3$ and a demanding energy convergence criterion of 10$^{-8}$~hartree ($2.72 \times 10^{-7}$~eV).
To integrate over the Brillouin zone, we used the tetrahedral method.
%
%
Moreover, we determined the values of excess electrons using the electronic population analysis introduced by Mulliken~\cite{mulliken_electronic_1955}.
The crystal-structure drawings we prepared with VESTA~\cite{momma_vesta_2008}.

\section{Results and discussion}

%
Although the very fact of the occurrence of perpendicular magnetic anisotropy in FeCo films of about four atomic monolayers thick is well known~\cite{shiota_voltage-assisted_2009}, the origin of this effect is not yet fully understood.
The difficulty in comprehension comes from the large number of factors that affect the magnetic anisotropy.
%
%
Among the most important of these are the composition, thickness, structural phase, and ordering parameter of the magnetic layer, growth directions of the substrate and magnetic layer, matching of the substrate and magnetic layer, structural deformation of the magnetic layer, the quality of the interfaces (their roughness and diffusion level), and temperature~\cite{lorenz_magnetic_1996}.

\subsection{Experiment}

%
\begin{figure}[!t]
\centering
\includegraphics[trim = -6 4 0 2,clip,width=0.96\columnwidth]{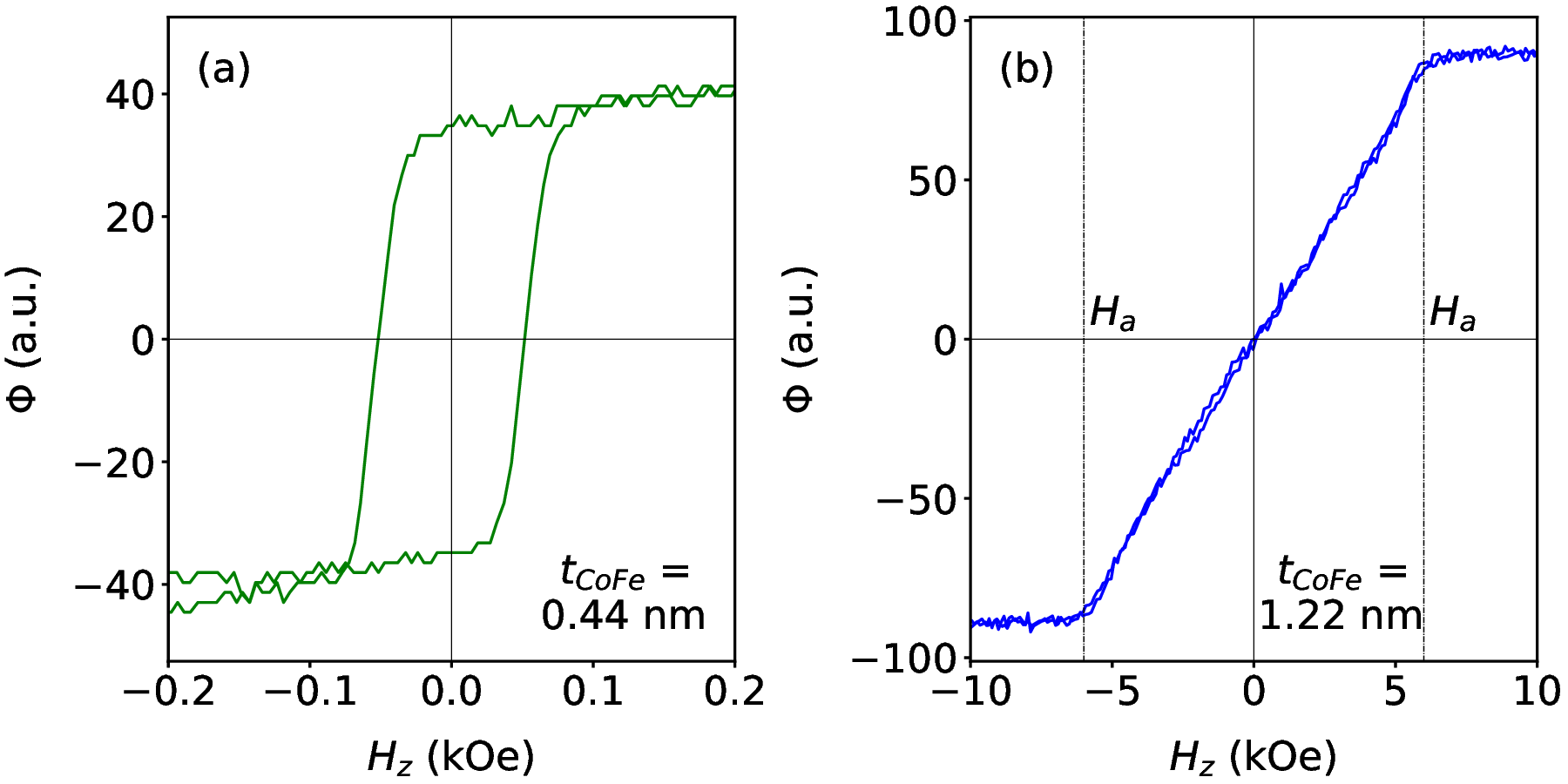}
\includegraphics[trim = 5 0 60 30,clip,width=0.99\columnwidth]{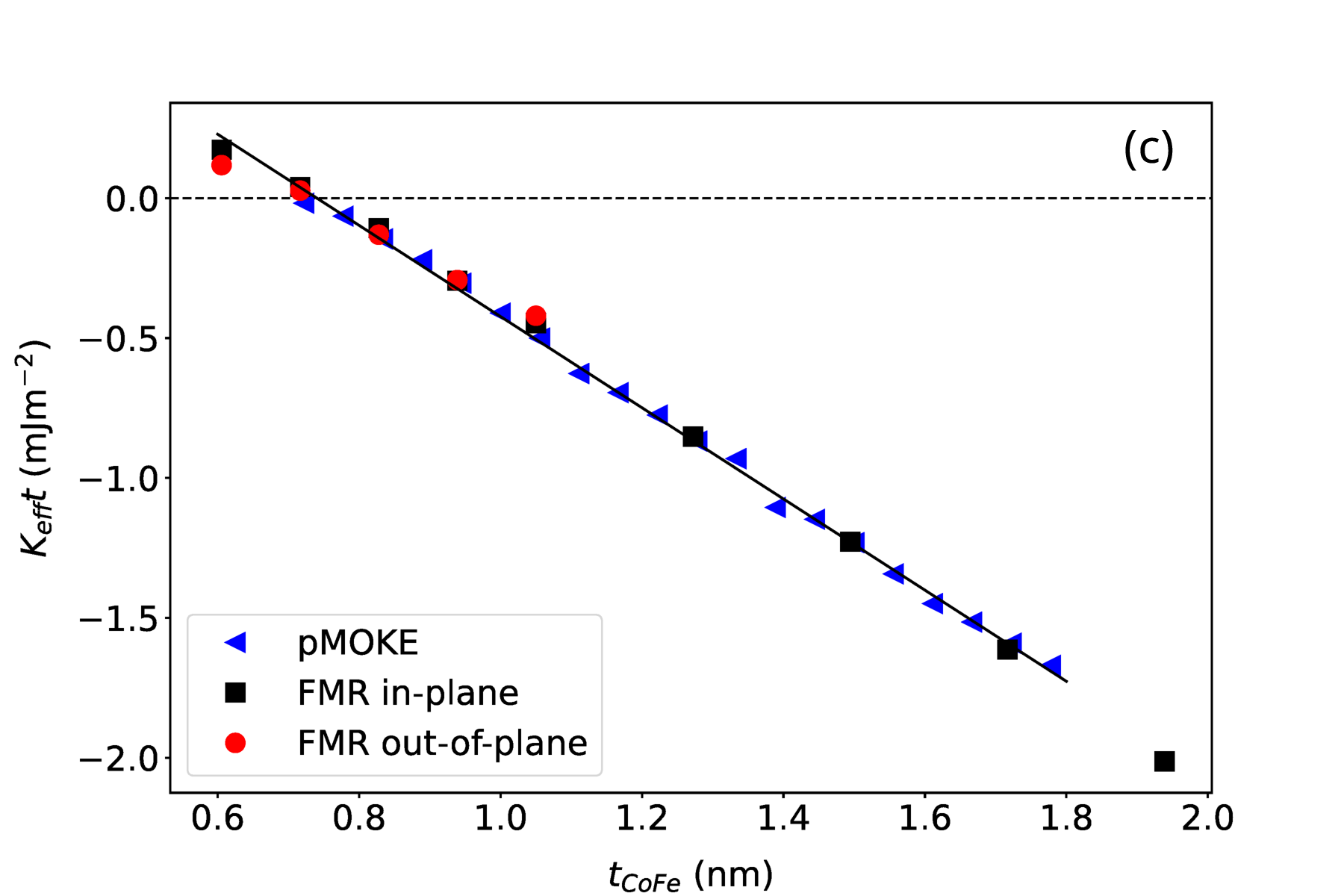}
\caption{
(a, b) The Kerr rotation $\phi$ as a function of perpendicular magnetic field $H_z$ (hysteresis loops) measured by the polar magneto-optical Kerr effect (pMOKE) for Fe$_{0.75}$Co$_{0.25}$ ultrathin films in Ti~3~nm\,/\,Au~60~nm\,/\,Fe$_{0.75}$Co$_{0.25}$\,/\,Au~3~nm samples. 
Results for \feco{} film thicknesses of (a) 0.44~nm and (b) 1.22~nm -- on the two sides of the spin-reorientation transition.
On panel (b) we indicated the anisotropy field ($H_{\rm{a}}$).
(c) The dependence of the product of effective magnetic anisotropy constant and film thickness ($K_{\rm{eff}}\,t$) on the Fe$_{0.75}$Co$_{0.25}$ film thickness,
measured with a ferromagnetic resonance spectrometer based on vector network analyzer (VNA-FMR) in out-of-plane and in-plane configurations
and carried out by the pMOKE method.
}
\label{fig:K_eff} 
\end{figure}

%
Perpendicular magnetic anisotropy (PMA) has been previously observed in Fe$_{0.8}$Co$_{0.2}$ layers sandwiched between Au(001) and MgO(001) layers, for FeCo film thickness of four atomic monolayers or less (<~0.55~nm)~\cite{shiota_voltage-assisted_2009}.
In that study, the MgO layer was chosen because it can serve as a spacer in a magnetic tunnel junction (FeCo/MgO/Fe)~\cite{shiota_induction_2012}.
In this work, we consider an ultrathin \feco{} layer sandwiched between two Au layers, see Fig.~\ref{fig:expt_heterostructure}.
Unlike in the aforementioned paper~\cite{shiota_voltage-assisted_2009}, we did not anneal our sample, therefore the structure is most-likely polycrystalline.

%
In the function of the effective magnetic anisotropy constant ($K_{\rm{eff}}\,t$) on the FeCo film thickness ($t$), see Fig.~\ref{fig:K_eff}, we observe a linear dependence, similar to that reported for Au/FeCo/MgO system~\cite{shiota_voltage-assisted_2009}.
However, in our case, the spin-reorientation transition, from out-of-plane to in-plane anisotropy, occurs at a film thickness of about five atomic monolayers (0.74~nm), not four (0.55~nm)~\cite{shiota_voltage-assisted_2009}.
%
%
In Fig.~\ref{fig:K_eff}, we show also the representative hysteresis loops measured for \feco{} film thicknesses below and above the spin-reorientation transition: at 0.44 and 1.22~nm.

%
\begin{table}[!ht]
    \centering
    \begin{tabular}{lccccc}
\hline
\hline
Method & System                                  &  $K_{\rm{V}}$    &  $2K_{\rm{S}}$ & Reference\\
       &                            & (MJ\,m$^{-3}$)   &  ($\mu$J\,m$^{-2}$) &                \\
\hline
expt. &   Au/Fe/MgO                               &  --       & 580 & Shiota~\cite{shiota_voltage-assisted_2009}\\
expt. &   Au/Fe$_{0.8}$Co$_{0.2}$/MgO            &  --       & 650  & Shiota~\cite{shiota_voltage-assisted_2009}\\
expt. &   [Fe$_{0.3}$Co$_{0.7}$/Au]$_{\times{}20}$            &  -0.80       & 700  & Wang~\cite{wang_magnetic_2011}\\
expt. &   Au/Fe$_{0.5}$Co$_{0.5}$/Au            &  ~0.69       & 650  & Żuberek~\cite{zuberek_magnetostriction_2000}\\
expt. &  Au/\feco{}/Au                     & -1.60     & 1200 & this work \\
DFT  &  Au/\feco{}/Au                & -1.36     & 3.3 & this work  \\
\hline
\hline
    \end{tabular}
    \caption{Volume magnetic anisotropy ($K_{\rm{V}}$) and surface magnetic anisotropy ($2K_{\rm{S}}$) determined from thickness dependence of effective magnetic anisotropy ($K_{\rm{eff}}$) with the formula: $K_{\rm{eff}}\,t = K_{\rm{V}}\,t + 2K_{\rm{S}}$ for ultrathin magnetic films deposited on Au.
    (In DFT calculations, to mimic the effect of room temperature, we reduced the spin magnetic moment of the systems to 80\% of the equilibrium value.)
    }
    \label{tab:kv_ks}
\end{table}

%
By convention, the linear range of the $K_{\rm{eff}}\,t\,(t)$
relation is interpreted as a combination of volume and surface contributions ($K_{\rm{V}}$ and $2K_{\rm{S}}$), according to the formula: $K_{\rm{eff}}\,t =  K_{\rm{V}}\,t +  2K_{\rm{S}}$~\cite{zuberek_magnetostriction_2000,shiota_voltage-assisted_2009}.
However reasonable this approach is for thicker films, for a thickness of several atomic monolayers, separating the film into a central volume part and two interface regions can be troublesome.
Nevertheless, a conventional approach is adopted in order to compare the heterostructure with systems that have been previously studied, prior to the presentation of a more nuanced interpretation.
For our multilayer, interpolation of the linear part of the $K_{\rm{eff}}\,t\,(t)$ leads to $K_{\rm{V}}$ of -1.6~MJ\,m$^{-3}$ and $2K_{\rm{S}}$ of 1200~$\mu$J\,m$^{-2}$.
The values differ significantly from results obtained for similar systems, see~Table~\ref{tab:kv_ks}.
Of particular note is the extremely low $2K_{\rm{S}}$ value determined from DFT calculations of 3.3~$\mu$J\,m$^{-2}$.
Since we observe large differences in $2K_{\rm{S}}$ between different experimental samples, we expect that the discrepancy between experimental and DFT results is also due to a difference in the microstructure of these systems. 
Namely, the layers we consider in the calculations are infinitely periodic, and thus our calculations do not describe the shape anisotropy present in the experimental samples. 
In addition, a description of the chemical disorder of the Fe-Co alloy using the virtual crystal approximation (VCA), which assumes the presence of a virtual atom of intermediate atomic number in place of a proportional concentration of Fe and Co atoms, may also lower $2K_{\rm{S}}$.

%
In summary, polycrystalline FeCo layers in Au/\feco{}/Au heterostructure show perpendicular magnetic anisotropy for thicknesses of five atomic monolayers, similar to analogous systems considered by other groups.
Our FeCo layer shows an unusually high value of surface anisotropy.
The small thickness of the layer with perpendicular magnetic anisotropy and the difficulty in unambiguously interpreting the contributions to magnetic anisotropy indicate the need for further analysis, for example using density functional theory (DFT).

%
\subsection{DFT calculations}

%
\subsubsection{Au$_\mathit{9}$/(Fe$_\mathit{0.75}$Co$_\mathit{0.25}$)$_\mathit{n}$/Au$_\mathit{9}$\, heterostructures}

%
\begin{figure}[!ht]
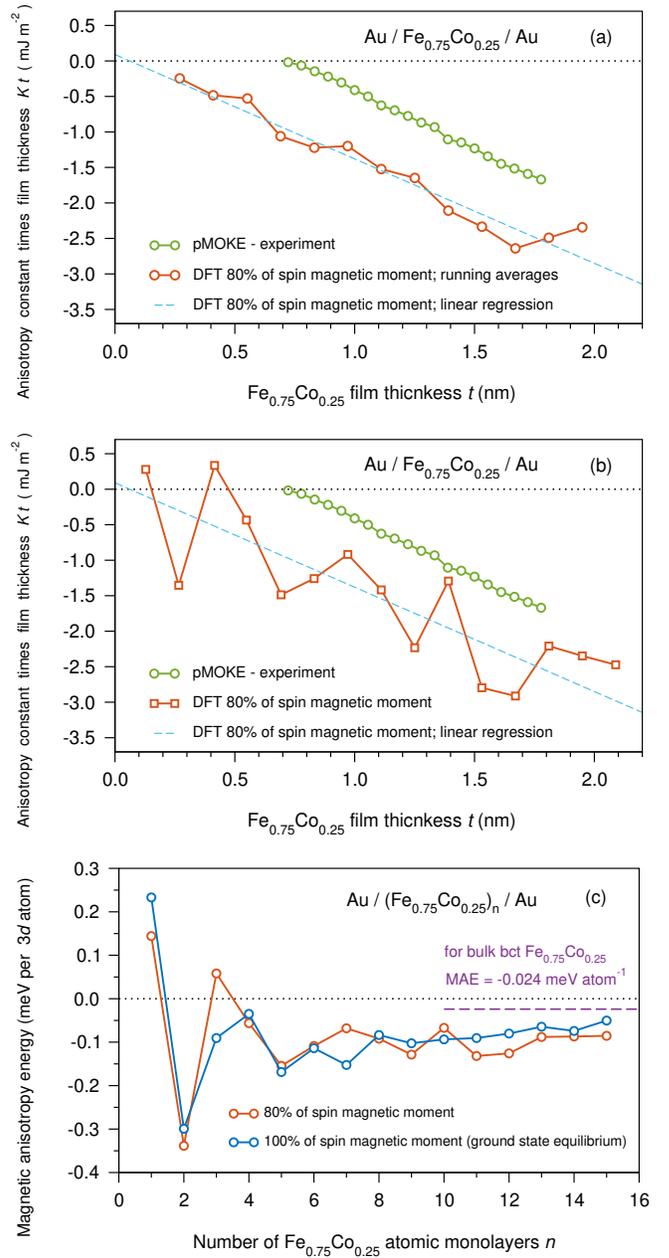

\centering
\includegraphics[clip,width=0.47\textwidth]{K_times_t_vs_t_expt_vs_ms08_averages.eps}\\
\vspace{3mm}
\includegraphics[clip,width=0.47\textwidth]{K_times_t_vs_t_ms08.eps}\\
\vspace{3mm}
\hspace{1.5mm}\includegraphics[clip,width=0.47\textwidth]{MAE_vs_n.eps}
\caption{
(a,b) Magnetic anisotropy constant times film thickness ($K\,t$) measured and calculated for Au/(\feco{})/Au heterostructures. 
The DFT results for Au$_\mathrm{9}$/(\feco{})$_{\mathrm{n}}$/Au$_\mathrm{9}$ heterostructures presented in panel (a) we obtained with reducing magnetic moment by 20\% and using running averages of three data-points of consecutive FeCo film thicknesses.
The DFT results shown in panel (b) are the same as the DFT results in panel (a) but we did not use averaging.
(c) Magnetic anisotropy energy (MAE) calculated as a function of the number of FeCo atomic monolayers $n$.
The DFT calculations were performed with the FPLO18 code using the PBE functional.
The virtual crystal approximation was applied to model the alloy in the \feco{} layer and bulk.
\label{fig:Kt_vs_t_expt.calc.}
}
\end{figure}

%
Figure~\ref{fig:Kt_vs_t_expt.calc.}(a) shows the calculated product of the magnetic anisotropy constant and film thickness ($K\,t$) as a function of the thickness, compared with the corresponding experimental values.
Since the measurements were performed at room temperature, in the calculations we set the value of the spin magnetic moment as 80\% of the nominal (equilibrium) value determined at 0~K to mimic the decrease in magnetic moment with temperature.
In addition, in order to account for the imperfections of film thickness present in the experimental systems, for calculation results we determined running averages of three consecutive thicknesses (e.g., $K\,t$ for 5 FeCo atomic monolayers is an average of the results for 4, 5 and 6 monolayers). 
Although the directions of the slope of the computed and measured slope are similar, the result of the computation is shifted downward compared to the experiment.
Part of the difference between computational and experimental results is accounted for by missing model elements, such as diffusion at interfaces, grain formation, dispersion of grain orientation, and chemical disorder in the FeCo alloy.

%
Figure~\ref{fig:Kt_vs_t_expt.calc.}(b) shows the basic computational results without averaging.
Although large deviations of individual data points from the averaged trajectory may appear to be caused by computational inaccuracy, such variations are typical of the dependence of magnetic anisotropy on layer thickness calculated in terms of the thickness of single atomic monolayers.~\cite{cinal_magnetocrystalline_1994,przybylski_oscillatory_2012, marciniak_l10_2023,marciniak_magnetic_2024}.
Figure~\ref{fig:Kt_vs_t_expt.calc.}(c) shows yet a third version of the discussed relationship, expressed without multiplying the anisotropy constant $K$ (MAE) by the thickness, and with the thickness given in the number of atomic monolayers instead of nanometers.
%
%
The MAE($n$) plots we obtained for 80\% and 100\% equilibrium magnetic moment are similar, see~Fig.~\ref{fig:Kt_vs_t_expt.calc.}(c).
Nevertheless, in some cases, such as for three monolayers, reducing the magnetic moment by 20\% significantly affect the MAE value.
Similar strong MAE dependence on magnetic moment (and indirectly on temperature) has been observed for some bulk materials, e.g. (Fe,Co)$_2$B alloys~\cite{edstrom_magnetic_2015}.
We have observed positive values of the magnetic anisotropy for monolayers of one and three atoms.
The MAE value calculated here for a single monolayer of FeCo between Au layers (0.23~meV\,atom$^{-1}$) is much lower than the MAE value (calculated with the use of DFT) for a freestanding monolayer of Fe (0.58~meV\,atom$^{-1}$), 
and ranks between bulk L1$_0$ phases of FeNi (calculated low MAE value of 0.024~meV\,atom$^{-1}$) and of FePt (calculated high MAE value of 1.38~meV\,atom$^{-1}$)~\cite{marciniak_l10_2023, marciniak_magnetic_2024}.
The magnetic anisotropy values calculated here for thicknesses other than one and three monolayers are negative, indicating in-plane anisotropy.
However, we observe a slow increase in anisotropy with layer thickness, 
which in the limit of thick layers converges to a value of -0.024~meV\,atom$^{-1}$ (-0.33~MJ\,m$^{-3}$), as determined by us for bulk body centered tetragonal (bct) \feco{} with a fixed lattice parameter $a$ of 2.88~\AA{} and an optimized lattice parameter $c$ of 2.805~\AA{}.
(The results for bulk bct \feco{} will be discussed later in this work.)

%
\begin{figure}[!t]
\centering
\includegraphics[clip,width=0.99\columnwidth]{E_minus_Eau_minus_Efeco_vs_n_in_eV.eps}\\
\vspace{2mm}
\includegraphics[clip,width=0.99\columnwidth]{c_vs_n.eps}
\caption{
(a) Difference between the total energy of the Au$_\mathrm{9}$/(\feco{})$_{\mathrm{n}}$/Au$_\mathrm{9}$ heterostructure and the energies of its bulk constituents: Au and \feco{}.
(b) Lattice parameter $c$ of \feco{} layers optimized with parameter $a$ fixed at the bulk value of $a_\mathrm{fcc Au}/\sqrt{2}$ equal to 2.88~\AA{}.
In the limit of thick films, the lattice parameter $c$ converges to 2.805~\AA{} calculated for bulk \feco{} with $a$ of 2.88~\AA{} (body centered tetragonal phase).
Calculations were carried out with the FPLO18 code using the PBE functional.
\label{fig:dE_and_c_vs_n}
}
\end{figure}
%
%
Figure~\ref{fig:dE_and_c_vs_n}(a) shows the energy of the heterostructure in relation to the energy of its bulk components.
The energy difference decreases exponentially as the thickness of the FeCo layer increases.
The observed increase indicates a decrease in the stability of the thinnest FeCo films.
This means a tendency towards island formation, structural deformation, structural reconstruction, and structural phase transition~\cite{lorenz_magnetic_1996, ravensburg_boundary-induced_2024}.
%
%
Figure~\ref{fig:dE_and_c_vs_n}(b) shows the step changes in the lattice parameter $c$ as the thickness of the magnetic layer decreases.
The lattice parameter $c$ of FeCo layers with thicknesses above about eight atomic monolayers is slowly increasing.
In the limit of large thicknesses, the lattice parameter $c$ of bct \feco{} converges to 2.805~\AA{} (calculated for fixed $a$ of 2.88~\AA{}).

%
\begin{figure}[!t]
\centering
\includegraphics[clip,width=0.99\columnwidth]{ms_vs_n.eps}
\caption{
Spin magnetic moment calculated for \feco{} layers in Au$_\mathrm{9}$/(\feco{})$_{\mathrm{n}}$/Au$_\mathrm{9}$ heterostructures.
Calculations were carried out with the FPLO18 code using the PBE functional.
\label{fig:ms_vs_n}
}
\end{figure}
The changes at the electronic and structural levels lead to changes in magnetic properties, see Fig.~\ref{fig:ms_vs_n}.
For large thicknesses, the spin magnetic moment tends toward the value of the bulk bct \feco{} of 2.35~$\mu_{\rm{B}}$\,atom$^{-1}$.
As the thickness of the layer decreases, the spin magnetic moment increases.
While for the thinnest layers with a thickness of a few monolayers, it also changes stepwise.
A similar exponential dependence of magnetic moment as a function of thickness has been observed experimentally for ultrathin Fe films~\cite{beltran_interfacial_2012}.
Significant changes in the magnetic anisotropy of the thinnest layers correlate with changes in their magnetic moment, i.e. the mutual shift of the valence band spin channels~\cite{edstrom_magnetic_2015}.

%
\begin{figure}[!t]
\centering
\includegraphics[clip,width=0.99\columnwidth]{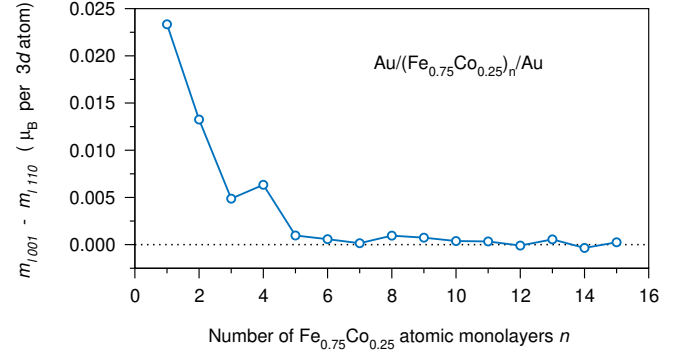}
\caption{
Orbital magnetic moment difference calculated for \feco{} layers in Au$_\mathrm{9}$/(\feco{})$_{\mathrm{n}}$/Au$_\mathrm{9}$ heterostructures.
Calculations were carried out with the FPLO18 code using the PBE functional.
\label{fig:dml_vs_n}
}
\end{figure}
The parameter associated with magnetic anisotropy is the difference in orbital magnetic moments determined in orthogonal directions~\cite{bruno_tight-binding_1989}.
For FeCo films of one to four atomic monolayers thick, the calculated values of the difference of orbital magnetic moment are significantly distinct from zero, see Fig.~\ref{fig:dml_vs_n}.

%
In summary, in this section we presented the properties of ultrathin FeCo films with a thickness of up to fifteen atomic monolayers. 
The majority of the continuous and smooth relationships observed for thicker films (e.g. MAE, $c$, $m_s$, $m_l$) transition to jittery below approximately ten monolayers.
This confirms that in this thickness regime, a change in layer thickness by a single monolayer can lead to significant qualitative changes in properties.
For \feco{} layers of one and three atomic monolayers thick, we predicted perpendicular magnetic anisotropy, while for two monolayers we predicted in-plane anisotropy.

%
\subsubsection{Au$_\mathit{9}$/(Fe$_\mathit{0.75}$Co$_\mathit{0.25}$)$_\mathit{15}$/Au$_\mathit{9}$\, heterostructure}
%
%
\begin{figure}[!ht]
\centering
\includegraphics[trim = 0 -20 0 0,clip,width=0.99\columnwidth]{excess_electrons_vs_distance.eps}
\includegraphics[clip,width=0.99\columnwidth]{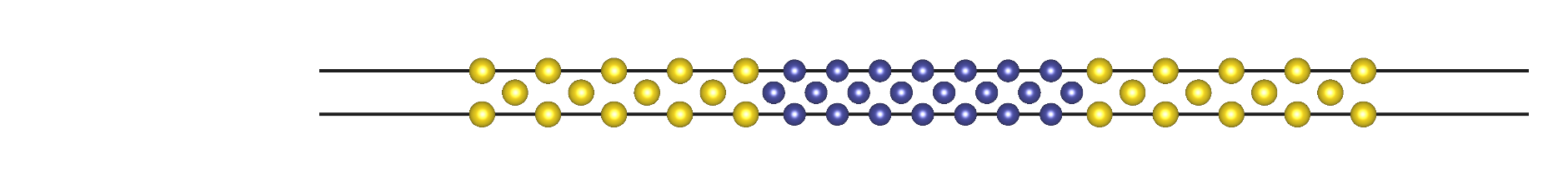}
\includegraphics[clip,width=0.99\columnwidth]{ms_vs_distance_v2.eps}
\caption{
Excess electrons (a) and spin magnetic moments (b) calculated for Au$_\mathrm{9}$/(Fe$_{\mathrm{0.75}}$Co$_{\mathrm{0.25}}$)$_{\mathrm{15}}$/Au$_\mathrm{9}$ heterostructure along the direction perpendicular to the film plane.
The positions of the atoms were expressed as the distance from the center of the FeCo layer.
The unit cell of the heterostructure is shown in the scale of the plot.
Calculations were carried out with the FPLO18 code using the PBE functional.
\label{fig:ee_ms_feco15}
}
\end{figure}
Following a detailed discussion on the FeCo layer thickness dependencies, the results for the selected 
Au$_\mathrm{9}$/(FeCo)$_{\mathrm{15}}$/Au$_\mathrm{9}$ 
heterostructure of fifteen atomic monolayers of FeCo (\mbox{$\sim2.1$}~nm) are now to be presented. 
Figure~\ref{fig:ee_ms_feco15} shows the excess electrons and spin magnetic moments calculated on atomic sites along the direction perpendicular to the film plane.
For excess electrons, the observed deviations range from about -0.05 to +0.05~$e$, which entail changes in the values of the magnetic moments.
The spin magnetic moments in the center of the \feco{} layer are close to the value for bulk bct \feco{} of 2.35~$\mu_{\rm{B}}$\,atom$^{-1}$,
while the first atomic monolayer at the interface has an elevated moment of 2.55~$\mu_{\rm{B}}$\,atom$^{-1}$.
This explains the increase in average magnetic moment with a decrease in layer thickness -- a consequence of the decrease in volume contribution, see Fig.~\ref{fig:ms_vs_n}.
Moreover, we see the induction of small magnetic moments in the two Au monolayers closest to the interface, being 0.015 and -0.006~$\mu_{\rm{B}}$\,atom$^{-1}$, which we also detected experimentally in our previous works~\cite{swindells_magnetic_2022, swindells_proximity-induced_2021}.

%
The biggest variation in calculated properties we observe at interfaces and surfaces; for about the four closest atomic monolayers bordering discontinuities.
This means that a layer with a thickness of less than about ten atomic monolayers consists of the atoms only with properties significantly deviated from those of analogous atoms in bulk systems, which help explain the unusual characteristics of the thinnest layers.

%
\subsubsection{Body centered tetragonal Fe$_\mathit{0.75}$Co$_\mathit{0.25}$ bulk }

%
\begin{figure}[!ht]
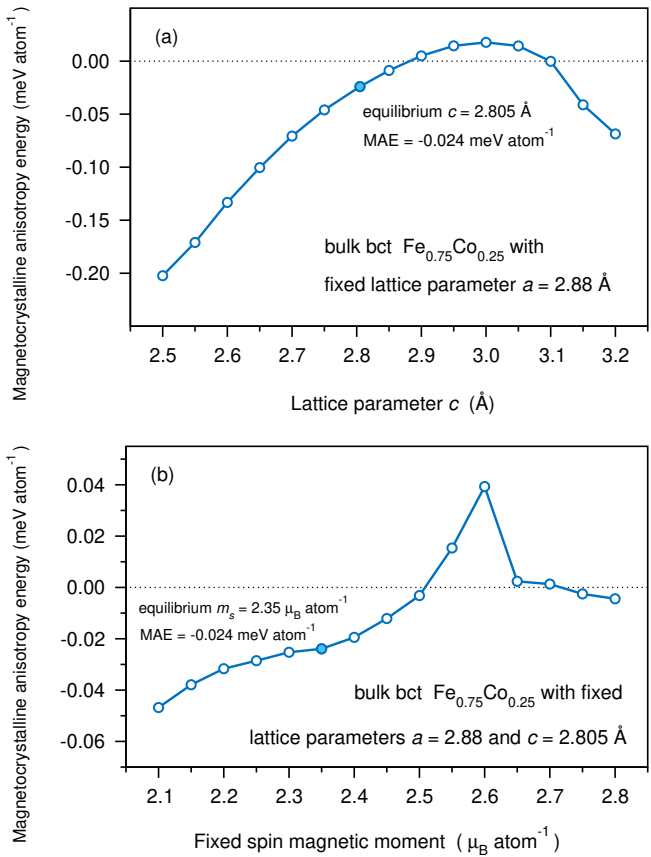

\centering
\includegraphics[trim = 0 -40 0 0,clip,width=0.99\columnwidth]{feco_bulk_MAE_vs_c_afixed2.88.eps}
\includegraphics[trim = 0 0 0 0,clip,width=0.99\columnwidth]
{feco_bulk_MAE_vs_fsm.eps}
\caption{
Magnetocrystalline anisotropy energy calculated for bulk body centered tetragonal (bct) \feco{} as a function of lattice parameter $c$ (panel a) and fixed spin magnetic moment (panel b).
Full circles indicate equilibrium values.
To model the \feco{} alloy, we applied the virtual crystal approximation.
Calculations were carried out with the FPLO18 code using the PBE functional.
\label{fig:MAE_vs_c_ms}
}
\end{figure}

As we have shown, the average values of the lattice parameter $c$ and the spin magnetic moment ($m_s$) of the \feco{} layer change as the thickness of the layer changes, see Figs.~\ref{fig:dE_and_c_vs_n}(b) and \ref{fig:ms_vs_n}.
To assess to what extent changes in $c$ and $m_s$ affect the value of the magnetic anisotropy energy of the layer, see Fig.~\ref{fig:Kt_vs_t_expt.calc.}(c), we performed additional calculations for a body centered tetragonal \feco{} bulk alloy, see Fig.~\ref{fig:MAE_vs_c_ms}.
%
%
For bulk bct \feco{} with a lattice parameter $a$ fixed at 2.88~\AA{} (epitaxial for fcc Au(001) substrate) and the corresponding optimized lattice parameter $c$ equal to 2.805~\AA{}, the MAE is -0.024~meV\,atom$^{-1}$ (-0.33~MJ\,m$^{-3}$); a value we have already referred to.
During compression ($c < 2.805$~\AA{}), MAE decreases.
In tension ($c > 2.805$~\AA{}) MAE increases; for $c = a = 2.88$~\AA{} (cubic structure) MAE passes through zero and remains positive (uniaxial magnetic anisotropy) up to $c = 3.1$~\AA{}, where it again passes through zero, this time back to in-plane anisotropy.
(The above pattern for \feco{} is known from the map of FeCo alloys with tetragonal deformation~\cite{burkert_giant_2004}.)
In the \feco{} layer with a thickness of 15 monolayers, the average value of the lattice parameter $c$ is about 2.788~\AA{}, which is slightly lower than the value of 2.805~\AA{} for bulk material, see Fig.~\ref{fig:dE_and_c_vs_n}(b).
With decreasing thickness of the FeCo layer, the value of the $c$ decreases steadily and for 8 monolayers is 2.781~\AA{}.
This reduction in $c$ correlates with a decrease in the value of MAE, see Fig.~\ref{fig:MAE_vs_c_ms}, and a similar trend is observed for the bulk \feco{}.
Below the thickness of 8 monolayers, the extreme values of $c$ are 2.569 and 2.781~\AA{}, for 2 and 3 monolayers, respectively.
The exceptionally low $c$ for the 2 \feco{} monolayers correlates with the low MAE value for both the \feco{} layer (-0.30~meV\,atom$^{-1}$) and the bulk material (-0.16~meV\,atom$^{-1}$).
Since, one MAE value is twice as large as the other, additional factors affecting the MAE of ultrathin layers come into play.

%
The second parameter on which the MAE strongly depends is the spin magnetic moment ($m_s$)~\cite{edstrom_magnetic_2015,marciniak_l10_2023,marciniak_magnetic_2024}. 
For bulk bct \feco{} with optimized $c$ equal to 2.805~\AA{}, the equilibrium $m_s$ is 2.35~$\mu_B$\,atom$^{-1}$.
The dependence of MAE on $m_s$, shown in Fig.~\ref{fig:MAE_vs_c_ms}(b), indicates that for the bulk bct \feco{} the MAE increases with $m_s$ until $m_s$ equals about 2.6~$\mu_B$\,atom$^{-1}$, where appears a discontinuity.
Above 2.5~$\mu_B$\,atom$^{-1}$, the MAE changes sign and the magnetic anisotropy changes to uniaxial.
Since a $m_s$ value of more than 2.5~$\mu_B$\,atom$^{-1}$ is observed on the outer atomic monolayer of the \feco{} layers, see~Fig.~\ref{fig:ee_ms_feco15}(b$_1$), we would also expect elevated MAE values for ultrathin films, especially with thicknesses from 1 to 5 monolayers.
Such dependence of MAE on $m_s$ may occur for the layers under consideration, but since their MAE is inevitably affected by the observed large changes in the lattice parameter $c$, we cannot conclusively resolve this.

%
In summary, changes in the magnetic anisotropy properties of ultrathin films can be probed from the MAE relationships determined for bulk material.
However, since the magnetic moments of the ultrathin layer are not homogeneous (and for the outermost monolayers change by leaps and bounds) to fully understand the dependence of MAE on layer thickness, an analysis based on a single average value of the magnetic moment is not sufficient.
Moreover, the above analysis further confirms that ultrathin layers with a thickness of less than about 10 atomic monolayers are significantly different from thicker layers. 
Below 10 monolayers, a thickness change by a single atomic monolayer can lead to significant qualitative changes in properties.
These changes are clearly visible in computational results for perfect atomically sharp interfaces, but they average out over thickness imperfections in experimental measurements.

\section{Summary and Conclusions}

In this paper, we have presented the results of experimental and computational studies for the Au/\feco{}/Au heterostructure with FeCo film thickness up to 2.1~nm (15 atomic monolayers).
Measurements showed the presence of perpendicular magnetic anisotropy in polycrystalline FeCo layers with a thickness of five atomic monolayers and fewer.
DFT calculations predicted step changes in magnetic anisotropy values for layers with thicknesses of several atomic monolayers, and anticipated perpendicular magnetic anisotropy for FeCo films with thicknesses of one and three atomic monolayers.
The abrupt changes in the anisotropy of thinnest layers are due to charge transfer in the monolayers closest to the interface and the followed changes in the magnetic moments.
Magnetic layers with a thickness of less than ten atomic monolayers are significantly different from the parent bulk material, as they are composed of two near-interface regions, and no longer contain a central volume part.
Systematic DFT calculations of the properties of magnetic heterostructures as a function of magnetic film thickness can help design magnetic junctions for spintronics applications.
This study highlights the potential of ultrathin FeCo layers for spintronic applications, driven by their robust perpendicular magnetic anisotropy (PMA) and tunable magnetic properties. The combination of pMOKE experiments and DFT simulations provides a comprehensive understanding of how thickness and interfacial effects influence their behavior.

\section{Acknowledgments}
We acknowledge the financial support of the National Science Centre Poland under the decision DEC-2018/30/E/ST3/00267 (SONATA-BIS 8). 
Part of the computations were performed on resources provided by the Poznan Supercomputing and Networking Center (PSNC). 
We thank P. Leśniak and D. Depcik for compiling the scientific software and administration of the computing cluster at the Institute of Molecular Physics, Polish Academy of Sciences.
We also thank Z. Śniadecki for his substantive comments on the manuscript.

\end{sloppypar} 

\bibliography{au_feco}

\end{document}